\pdfoutput=1

\documentclass[11pt,twoside,a4paper,cmspaper,final,collab]{cms-tdr}
\def\svnVersion{482963P}\def\cmsCernNoTag{CERN-EP-2018-121}\def\cmsCernDate{\today}\def\cmsMessage{Published in the Journal of High Energy Physics as \href{http://dx.doi.org/10.1007/JHEP11(2018)115}{\doi{10.1007/JHEP11(2018)115.}}}

\begin{document}\cmsNoteHeader{HIG-16-030}

\hyphenation{had-ron-i-za-tion}
\hyphenation{cal-or-i-me-ter}
\hyphenation{de-vices}
\RCS$HeadURL: svn+ssh://svn.cern.ch/reps/tdr2/papers/HIG-16-030/trunk/HIG-16-030.tex $
\RCS$Id: HIG-16-030.tex 476672 2018-10-01 10:25:12Z gbyu $
\providecommand{\cmsTable}[1]{\resizebox{\textwidth}{!}{#1}}
\providecommand{\NA}{\ensuremath{\text{---}}}
\providecommand{\CL}{CL\xspace}
\newlength{\cmsTabSkip}\setlength{\cmsTabSkip}{1.5ex}
\newcommand{\RNum}[1]{\uppercase\expandafter{\romannumeral #1\relax}}
\newcommand{\cb}{\ensuremath{\cPqc\cPaqb}\xspace}
\newcommand{\ch}{\ensuremath{\PH^+}\xspace}
\newcommand{\mh}{\ensuremath{M_{\ch}}\xspace}
\newcommand{\tchb}{\ensuremath{\cPqt \to \ch \cPqb}\xspace}
\newcommand{\twb}{\ensuremath{\cPqt \to \PW \cPqb}\xspace}
\newcommand{\hcb}{\ensuremath{\ch \to \cPqc \cPaqb}\xspace}
\newcommand{\reliso}{\ensuremath{I_\text{rel}}\xspace}

\newcommand\Tstrut{\rule{0pt}{2.6ex}}
\newcommand\Bstrut{\rule[-0.9ex]{0pt}{0pt}}
\newcommand{\TBstrut}{\Tstrut\Bstrut}

\newcommand{\ttbb}{\ensuremath{\ttbar\bbbar}\xspace}
\newcommand{\br}{\ensuremath{\mathcal{B}}\xspace}
\newcommand{\pzn}{\ensuremath{p_{z}^{\Pgn}}\xspace}

\cmsNoteHeader{HIG-16-030}
\title{Search for a charged Higgs boson decaying to charm and bottom quarks in proton-proton collisions at \texorpdfstring{$\sqrt{s} = 8\TeV$}{sqrt(s) = 8 TeV}}

\date{\today}

\abstract{
A search for charged Higgs boson decaying to a charm and a bottom quark ($\hcb$) is performed using 19.7\fbinv of pp collision data at $\sqrt{s} = 8\TeV$.  The production mechanism investigated in this search is $\ttbar$ pair production in which one top quark decays to a charged Higgs boson and a bottom quark and the other decays to a charged lepton, a neutrino, and a bottom quark. Charged Higgs boson decays to $\cb$ are searched for, resulting in a final state containing at least four jets, a charged lepton (muon or electron), and missing transverse momentum. A kinematic fit is performed to identify the pair of jets least likely to be the bottom quarks originating from direct top quark decays and the invariant mass of this pair is used as the final observable in the search. No evidence for the presence of a charged Higgs boson is observed and upper limits at 95\% confidence level of 0.8--0.5\% are set on the branching fraction $\br(\tchb)$, assuming $\br(\hcb) = 1.0$ and $\br(\tchb) + \br(\twb) = 1.0$,  for the charged Higgs boson mass range 90--150\GeV.
}

\hypersetup{
pdfauthor={CMS Collaboration},
pdftitle={Search for a charged Higgs boson decaying to charm and bottom quarks in proton-proton collisions at sqrt(s) = 8 TeV},
pdfsubject={CMS},
pdfkeywords={Charged Higgs, 2HDM type-Y, 2HDM flipped, top quark decay property}
}

\maketitle

\section{Introduction}

In 2012, a boson with a mass about 125\GeV was discovered at the CERN LHC~\cite{HiggsATLAS12,HiggsCMS12,HiggsCMSlong} with its properties subsequently shown~\cite{Khachatryan:2014jba,Aad:2015zhl,Aad:2015gba,Khachatryan:2016vau} to be consistent with those of the standard model (SM)~\cite{ref:sm1,ref:sm2,ref:sm4} Higgs boson~\cite{Higgs1,Higgs2,Higgs3,Higgs4,Higgs5,Higgs6}.
Although the last missing particle of the SM has been discovered, several questions remain, including the nature of dark matter~\cite{dark1,dark2}, and the origin of neutrino masses~\cite{pdg} inferred from the observation of neutrino oscillations~\cite{Fukuda:1998mi}.
Several hypotheses beyond the SM have been introduced and tested to answer these questions, and many of them include more than one Higgs doublet.
Models with two Higgs doublets, so-called two-Higgs-doublet model (2HDM)~\cite{Gunion:1989we,ref:2hdm1},  result in five Higgs bosons: two charged (\Hpm) and three neutral ($\PSA$, $\PH$, $\Ph$). In the 2HDM, the Higgs boson discovered at the LHC can be one of the CP-even neutral bosons ($\PH$ or $\Ph$). Unlike the SM, in general 2HDM allows flavour changing neutral current (FCNC) at tree level.  To suppress such tree level FCNC, all fermions with the same electric charge are required to couple to one Higgs doublet only~\cite{fcnc1,fcnc2}. The 2HDM is typically categorized into four different types: type-\RNum{1}, type-\RNum{2}, lepton-specific (type-\RNum{3}), and flipped (type-Y, also known as type-\RNum{4}), depending on the assignment of up/down-type quark and lepton couplings to each Higgs doublet.

We present a search for charged Higgs bosons. Hereafter, we refer to them as $\ch$, but charge conjugate states are always implied.  In the 2HDM, the mass of the charged Higgs boson ($\mh$) is an unconstrained parameter. Regardless of its mass, $\ch$ is expected to have a large coupling to the top quark unless a specific condition is being considered as in Refs.~\cite{ref:2hdmf1,ref:2hdmf2}.
If $\mh$ is smaller than the top quark mass, the so-called light charged Higgs boson scenario, the top quark can decay to  a $\ch$ and a $\cPqb$ quark, $\tchb$.
The LEP experiments~\cite{lep} excluded the mass of charged Higgs below 80 (72.5)\GeV for type-\RNum{2} (type-\RNum{1} for pseudo-scalar masses above 12\GeV) scenario at 95\% confidence level ($\CL$). In the presence of the $\PW$ boson resonance at a mass of 80.4\GeV, the light charged Higgs boson search range is typically set between the $\PW$ boson mass and the top quark mass.
Previous direct searches for a light $\ch$ in decays of a top quark have been performed at hadron collider experiments in following channels: $\ch \to \overline{\tau}\Pgn$~\cite{ref:cdf_tn,ref:d0_tn,ref:cdf_tn2,ref:cms_tn,ref:atlas_tn,ref:atlas_tn2,ref:atlas_tn3}, $\ch \to \cPqc\cPaqs$~\cite{ref:cdf_cs,ref:d0_cs,ref:atlas_cs,ref:cms_cs}, and $\ch \to \PW\PSA$~\cite{ref:cdf_wa}. No indication of a $\ch$ was observed and the best upper limits on the branching fraction of $\tchb$ were placed at $\mathcal{O}$(1\%).
The $\hcb$ process is the dominant decay channel in the type-Y 2HDM~\cite{ref:typeY,ref:typeY2,Akeroyd:2012yg}, and this signal could be a signature of models with more than two Higgs doublets~\cite{Grossman:1994jb,Akeroyd:1994ga}.
The search is performed assuming $\br(\hcb) = 1.0$ without any other model-dependent assumption.

\begin{figure}[htbp]
\centering
    \includegraphics[width=0.48\textwidth]{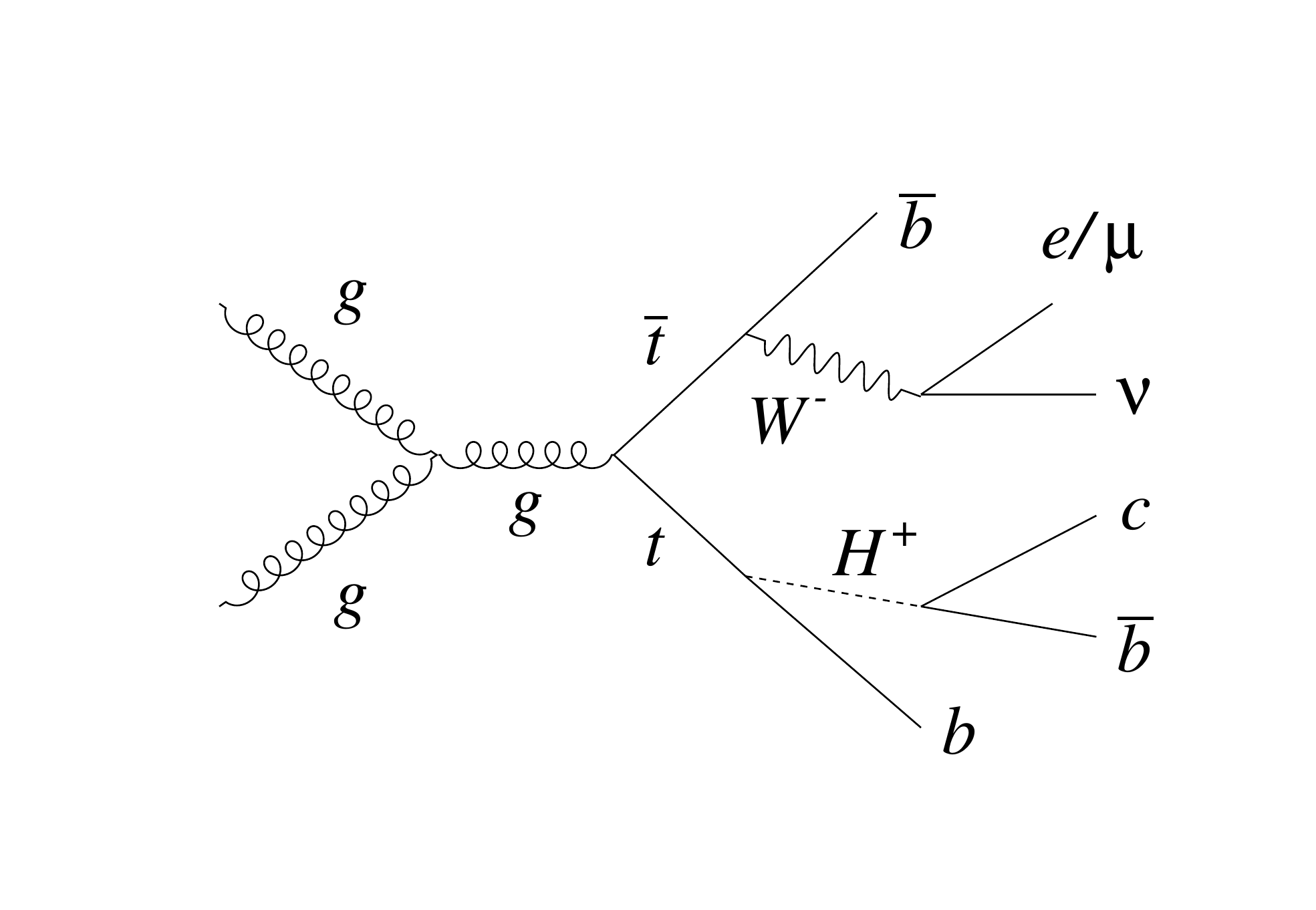}
    \includegraphics[width=0.48\textwidth]{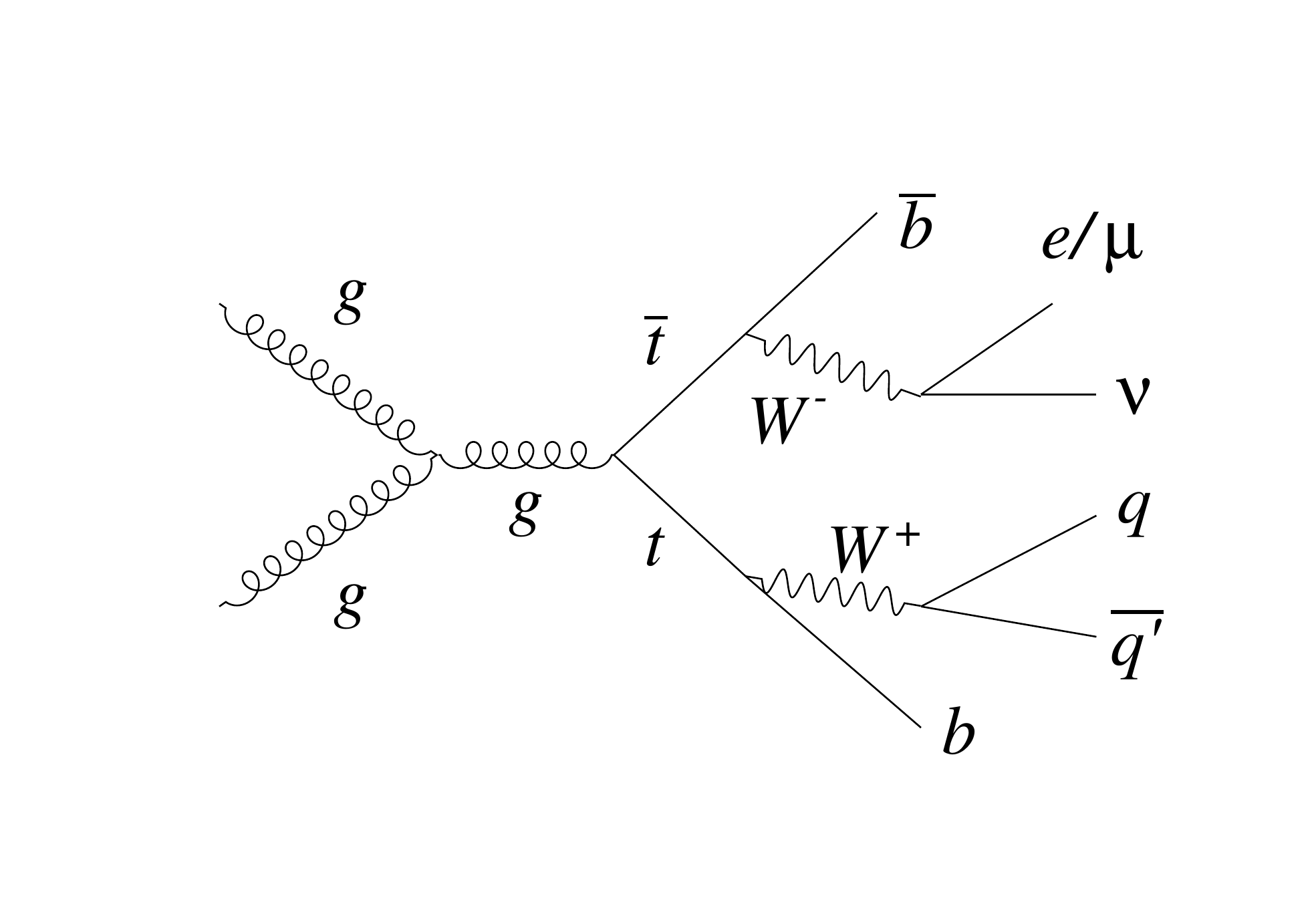}
\caption{Feynman diagrams of the $\ch$ production in top quark pair events (left) compared to the standard model production of $\ttbar$ in lepton+jets final states (right).  }\label{fig:diagram}
\end{figure}

The search uses $\ttbar$ events with a final state of at least four jets (at least two of which originate from $\cPqb$ quarks), a charged lepton (muon or electron), and missing transverse momentum. If a light $\ch$($\to \cb$) is produced in top quark decays, the $\ttbar$ event would have one more jet to be identified originating from $\cPqb$ quark due to the $\ch$ decays, as shown in Fig.~\ref{fig:diagram}.  A kinematic fit is performed to identify the pair of jets least likely to be the $\cPqb$ quarks originating from direct top quark decays. The invariant mass of this jet pair is used as the final observable in this search. The signal events are expected to peak at the charged Higgs boson mass. We assume $\br(\tchb) + \br(\twb) = 1.0$, which implies a lowering of the branching fraction of top quarks to $\PW\cPqb$ in presence of $\ch$ in top quark decays.

The main background for this search is SM $\ttbar$, including $\ttbar$ production in association with heavy-flavoured jets ($\ttbb$, $\ttbar\ccbar$).  Other considered backgrounds are single top production, multijet, $\PW/\PZ$+jets and diboson production, and $\ttbar$ production in association with an $\PH/\PZ/\PW$ boson.

\section{Event simulation and reconstruction with CMS detector}

Background samples of $\ttbar$, $\ttbar$+$\PW/\PZ$, and $\PW/\PZ$+jets are simulated at leading order (LO) using the $\MADGRAPH$ 5.1 generator~\cite{ref:MG} with the CTEQ6L1 parton distribution function (PDF) set~\cite{CTEQ6L}. The top quark mass is set to 172.5\GeV for simulating these samples.
The predicted $\ttbar$ production cross section is calculated with the \textsc{Top++} 2.0 program at the next-to-next-to-leading order (NNLO) in perturbative quantum chromodynamics (QCD), including soft-gluon resummation at the next-to-next-to-leading-log order (Ref.~\cite{topxsec1} and references therein), to be $\sigma_{\ttbar} = 252.9^{+6.4}_{- 8.6}\,\text{(scale)}\pm 11.7\,\text{(PDF+$\alpS$)}\unit{pb}$, where ``scale'' and ``PDF+$\alpS$'' refer to the uncertainties coming from the independent variation of the factorization and renormalization scales, and the variations in the PDF set and in the strong coupling constant $\alpS$, respectively,  following the PDF4LHC prescription with the MSTW2008 68\% \CL NNLO, CT10 NNLO and NNPDF2.3 5f FFN PDF sets (Refs.~\cite{Alekhin:2011sk,topxsec2} and references therein, and Refs.~\cite{topxsec3,topxsec4,topxsec5}).

The transverse momentum $\pt$ distribution of top quarks in simulated $\ttbar$ events is reweighted to match the $\pt$ distribution observed in collision data~\cite{Khachatryan:2015oqa}.
The simulated $\PW/\PZ$+jets samples are normalized to the NNLO cross section calculated with
 \FEWZ 3.1~\cite{fewz31,FEWZ2}, and $\ttbar$+$\PW / \PZ$ events are normalized to the next-to-leading order (NLO) cross section~\cite{ttwxsec,ttzxsec}.
Single top quark events are generated with the \POWHEG v1.0 generator~\cite{ref:powheg,powheg1,powheg2,powheg3} and the CTEQ6M PDF set~\cite{CTEQ6L}, and are normalized to the production cross section at NLO in QCD computed with HATHOR v2.1~\cite{hathor1,hathor2}.
Diboson ($\PW\PW / \PW\PZ / \PZ\PZ$) and $\ttbar\PH$ events are generated at LO using \PYTHIA v6.4~\cite{ref:pythia} and normalized to the NLO cross section calculated using \MCFM 6.6~\cite{mcfm} and the cross section given in Ref.~\cite{cern3}, respectively.

The charged Higgs boson signal events ($\ttbar \to \cPqb\ch\cPaqb\PWm \to \bbbar\cb\ell\Pgn$) are simulated using the \PYTHIA v6.4 and CTEQ6L1 PDF set for $\mh = 90$, 100, 110, 120, 130, 140, and 150\GeV. These samples are normalized to the SM $\ttbar$ cross section in lepton+jets channel. Consequently, in the assumption of  $\br(\hcb) = 1.0$ and $\br(\tchb) + \br(\twb) = 1.0$, a fit using templates of the SM $\ttbar$ and the $\ch$ signal determines the branching fraction of $\tchb$.

All generated samples are interfaced with \PYTHIA~v6.4 in order to simulate parton showering and hadronization, and then processed through the full simulation of the CMS detector based on \GEANTfour~\cite{geant4}.
The underlying event tune Z2*~\cite{tunez1,tunez2} is used. To ensure correct simulation of the number of additional interactions per bunch crossing (pileup), simulated events are mixed with multiple inelastic collision events and reweighted according to the distribution of the number of pileup interactions observed in data.

The central feature of the CMS apparatus is a superconducting solenoid of 6 m internal diameter,
providing a magnetic field of 3.8\unit{T}. Within the solenoid volume are a silicon pixel and strip tracker,
a lead tungstate crystal electromagnetic calorimeter (ECAL), and a brass and scintillator hadron calorimeter (HCAL),
each composed of a barrel and two endcap sections. Additional forward calorimetry complements the coverage provided
by the barrel and endcap detectors.  Muons are detected in gas-ionization chambers embedded in the steel flux-return yoke outside the solenoid.
A more detailed description of the CMS detector, together with a definition of the coordinate system used and the relevant kinematic variables, can be found in Ref.~\cite{CMS_detector}.

A particle-flow (PF) algorithm~\cite{CMS-PRF-14-001} aims to reconstruct and identify particle candidates with an optimized combination of information from various elements of the CMS detector. Muon momenta are obtained from the curvature of muon tracks. The energy of photons is obtained from the ECAL measurement, upon proper calibration of several instrumental effects as described in~\cite{Chatrchyan:2013dga,Khachatryan:2015iwa}.
The energy of electrons is determined from a combination of the electron momentum at the primary interaction vertex (PV) as determined by the tracker, the energy of the corresponding ECAL cluster, and the energy sum of all bremsstrahlung photons spatially compatible with originating from the electron track~\cite{Khachatryan:2015hwa}. The PV is the reconstructed vertex with the largest value of $\sum \pt^{2}$, the sum of squared transverse momenta of the charged particle tracks associated with the vertex. The energy of charged hadrons is determined from a combination of their momentum measured in the tracker and the matching ECAL and HCAL energy deposits.
Finally, the neutral hadrons are identified as HCAL energy clusters not linked to any charged hadron trajectory, or as ECAL and HCAL energy excesses with respect to the expected charged hadron energy deposit or photon.

Jets are reconstructed from all the PF candidates clustered using the anti-\kt algorithm~\cite{Cacciari:2008gp, Cacciari:2011ma} with a distance parameter of 0.5. The jet momentum is determined as the vectorial sum of all particle momenta in the jet, and corrected for effects of pileup within the same or nearby bunch crossings.
Jet energy scale corrections~\cite{Khachatryan:2016kdb,Chatrchyan:2011ds} are used to account for the nonlinear energy response of the calorimeters and other instrumental effects.
Additional selection criteria are applied to each event to remove spurious jet-like features originating from isolated noise patterns in certain HCAL regions.
The missing transverse momentum vector $\ptvecmiss$ is defined as the projection onto the plane perpendicular to the beam axis of the negative vector sum of the momenta of all reconstructed PF objects in an event. Its magnitude is referred to as $\ptmiss$.

\section{Event selection and yields}

Candidate signal events are selected using triggers~\cite{Khachatryan:2016bia} that require a single isolated muon (electron) with $\pt >  24$ (27)\GeV and pseudorapidity $\abs{\eta} < 2.1$ (2.5).
Further selection requirements are made offline.
Events with exactly one muon (electron) with $\pt > 26$\,(30)\GeV and $\abs{\eta} < 2.1$\,(2.5) are selected.
Lepton identification selections, including requirements of a good track quality and close distance with respect to the PV, are imposed on each lepton candidate.
Leptons must be isolated, satisfying relative isolation requirement $\reliso < 0.12$\,(0.1) for muons (electrons).  The $\reliso$ is defined as the pileup-corrected scalar $\pt$ sum around the lepton candidate's direction at the vertex divided by the lepton candidate $\pt$. The $\pt$ sum is calculated from  momenta of the reconstructed charged hadrons originating from the PV, neutral hadrons, and photons within a cone of $\DR=\sqrt{\smash[b]{(\Delta\eta)^2+(\Delta\phi)^2}}  < 0.4$ (0.3) for muons (electrons), where $\phi$ is the azimuthal opening angle (in radians). Events with any additional muons (electrons) satisfying $\pt > 10\,(20)$,  $\abs{\eta} < 2.5$, and $\reliso < 0.3$, are discarded.

The $\ptmiss$ is required to be larger than 20\GeV, and at least four jets are required to have $\pt > 30\GeV$ within the tracker coverage of $\abs{\eta} < 2.4$.
To identify jets originating from $\cPqb$ quarks, the combined secondary vertex tagging algorithm~\cite{ref:btag} is used. Selected jets are considered $\cPqb$-tagged if they satisfy the medium working point requirements of this algorithm.
This results in an efficiency of approximately 70\% for tagging a $\cPqb$ quark jet, and a mistag rate of 1\% for light quark and gluon jets. The probability of a $\cPqc$ jet to be tagged as a $\cPqb$ jet is about 20\%. Events with two or more $\cPqb$-tagged jets are selected.

The events selected using the above criteria are dominated by SM $\ttbar$ events ($\approx$92\%) based on the background simulation samples.
The observed event yields in events with two $\cPqb$-tagged jets are well described by the simulation, however, the events containing three or more $\cPqb$-tagged jets are more difficult to model.
In order to estimate the $\ttbar$ component in the three or more $\cPqb$-tagged jet event sample, we rely on the measurement of the $\ttbb$ cross section in Ref.~\cite{CMS:2014yxa}.   In this reference, the $\ttbb$ cross section is measured to be $0.36 \pm 0.08\stat \pm 0.1\syst$\unit{pb}. Comparing with the theoretical expectation of $0.23 \pm 0.05$\unit{pb}, we obtain a ratio between the measured and the expected $\ttbb$ production cross section of 1.56 $\pm$ 0.66. As the study used dilepton $\ttbar$ events of same generator with current $\ttbar$ simulation sample, in which both top quarks decay to $\PW\cPqb$ with $\PW \to \ell \Pgn$, the $\ch(\to \cb)$ contribution to this extra $\cPqb$ quark process is negligible. The events with only one extra $\cPqb$ jet ($\ttbar \cPqb \mathrm{j}$) is understood to come from the $\ttbb$ process with one $\cPqb$ jet missed. Consequently, the $\ttbb$ component in the simulated $\ttbar$ sample is estimated by requiring at least one additional jet originated from an extra $\cPqb$ quark based on generator information, then rescaled by the $\ttbb$ cross section ratio.

The multijet background is estimated following the method used in Ref.~\cite{ref:cms_cs}.
The shapes of the multijet background distributions are obtained from a nonisolated control region defined by $0.15 < \reliso < 0.3$ and  $\ptmiss > 20$\GeV, after subtraction of the estimated SM backgrounds.
In a QCD enhanced control region ($\ptmiss < 20\GeV$), a multiplicative scale factor used for the multijet background normalization is obtained from the nonisolated control region extrapolated to the isolated region.  The shape uncertainty is estimated from the multijet background samples obtained using the same method but with shifted nonisolated control regions, $0.2 < \reliso < 0.3$ (smaller statistics) and reversing $\reliso$ selection, $0.12 (\Pgm)/0.1 (\Pe) < \reliso < 0.3$ (larger statistics).  The normalization uncertainty is estimated by an average difference in the multijet background yields obtained from the shifted nonisolated control regions compared to the nominal multijet background, and its impact on the total SM backgrounds except the $\ttbar$ process (non-$\ttbar$) is calculated to be 10\% or less.

Event yields satisfying the selection criteria in the absence of a signal are summarized in Table~\ref{tab:yield}. The $\ttbar$ event yields are estimated after rescaling the $\ttbb$ component.
The number  of $\cPqb$-tagged jets ($\cPqb$ tags) indicated in Table~\ref{tab:yield} is the number of $\cPqb$ tags among the four jets with highest $\pt$ in the event, which are used in the $\ttbar$ reconstruction. Signal efficiency satisfying the selection criteria is 4--6\% depending on $\mh$.

\begin{table}[hptb]
\topcaption{Observed event yields and estimated backgrounds for the $\Pgm$+jets and $\Pe$+jets channels satisfying the event selection criteria. The number of $\cPqb$-tagged jets is the number of $\cPqb$ tags among the four jets with highest $\pt$ in the event. The first and second uncertainty shown corresponds to the statistical and systematic components, respectively.}\label{tab:yield}
\centering
\cmsTable{
\begin{tabular}{l| r@{\hskip1pt}l@{\hskip1pt}l | r@{\hskip1pt}l@{\hskip1pt}l | r@{\hskip1pt}l@{\hskip1pt}l | r@{\hskip1pt}l@{\hskip1pt}l  }
\hline
       & \multicolumn{6}{c|}{$\Pgm$+jets} & \multicolumn{6}{c}{$\Pe$+jets} \Tstrut\\
       & \multicolumn{3}{l}{2 $\cPqb$ tags} &  \multicolumn{3}{l|}{$\geq$3 $\cPqb$ tags} &  \multicolumn{3}{l}{2 $\cPqb$ tags} &  \multicolumn{3}{l}{$\geq$3 $\cPqb$ tags } \Bstrut \\
       \hline
               $\ttbar$ &  52821 &$\pm$ 67 &$\pm$ 5463 & 5060 &$\pm$ 21 &$\pm$ 586 & 44484 &$\pm$ 60  &$\pm$ 4682  & 4269 &$\pm$ 19 &$\pm$ 468 \Tstrut \\
               Single top &  2212 &$\pm$ 30 &$\pm$ 178 & 169 &$\pm$ 8 &$\pm$ 16 & 1882 &$\pm$ 28 &$\pm$ 161  & 147 &$\pm$ 8 &$\pm$ 13 \\
               $\ttbar$+$\PW/\PZ/\PH$  & 195 &$\pm$ 2  &$\pm$ 8 & 41 &$\pm$ 1 &$\pm$ 3 & 169 &$\pm$ 2  &$\pm$ 7 & 35 &$\pm$ 1 &$\pm$ 2\\
               $\PW/\PZ$+jets &   1305 &$\pm$ 127 &$\pm$ 157 & 13 &$\pm$ 7 &$\pm$ 13 & 1098 &$\pm$ 114 &$\pm$ 165  & 32 &$\pm$ 19 &$\pm$ 14\\
               $\PW\PW$/$\PW\PZ$/$\PZ\PZ$ & 62 &$\pm$ 2 &$\pm$ 7  & 5 &$\pm$ 1& $\pm$ 1 & 56 &$\pm$ 2 &$\pm$ 6  & 4 &$\pm$ 1 &$\pm$ 1\\
               Multijet &  497 &$\pm$ 15 &$\pm$ 15 & 190 &$\pm$ 19 &$\pm$ 23 & 996 &$\pm$ 31 &$\pm$ 58  & 178 &$\pm$ 17 &$\pm$ 20  \Bstrut \\[\cmsTabSkip]
Expected
&   57093 &\multicolumn{2}{@{\hskip1pt}l|}{$\pm$ 5470 (stat+syst)}
& 5477 & \multicolumn{2}{@{\hskip1pt}l|}{$\pm$ 588 (stat+syst)}
&  48683 &\multicolumn{2}{@{\hskip1pt}l|}{$\pm$ 4688 (stat+syst)}
& 4665 &\multicolumn{2}{@{\hskip1pt}l}{$\pm$ 470 (stat+syst)}  \Tstrut \\
Observed &  57593 &           &              & 5754 &           &            &  50542 &             &            & 4848 & & \Bstrut \\
\hline
\end{tabular}
}
\end{table}

\section{Reconstruction of \texorpdfstring{$\ttbar$}{ttbar} events}\label{rec_tt}

Top quark and $\PW$ boson masses are reconstructed relying on the knowledge of the momenta of their decay products.
However, the reconstructed mass is different from the true mass because the measured jet energy is corrected to the energy of a particle-level jet, not to the energy of the initial parton.
A correction is derived from the energy shift between a particle-level jet and the matched hard scattering parton within $\DR =$ 0.3, depending on its matched parton flavour ($\cPqb$, $\cPqc$, or light quarks) in the SM $\ttbar$ simulation sample. This correction is called  the top quark specific (TS) correction and is applied as a function of the $\pt$ and $\eta$ of the jet. The application of TS correction in the $\ttbar$ reconstruction have been used in several analyses~\cite{ref:cdf_cs,ref:atlas_cs,Abulencia:2005aj}. Using this correction increases the accuracy of the mass reconstruction for top quarks and $\ch/\PW$ boson decaying to dijet, resulting in a 7--9\% improvement in resolution.

The instrumental mass resolution is further improved using a kinematic fit.
The fit is used to fully reconstruct the $\ttbar$ system by assigning selected jets to the hadronic $\PW/\ch$ decays or $\cPqb$ quarks in $\ttbar$ decays. The function that is minimized in the fit is as follows:
\begin{linenomath}
\begin{multline}
\chi^{2}  = \sum_{\pzn\ \text{solutions}} \Biggl( \sum_{i=\ell,\ 4\text{jets}} \frac{(\pt^{i, \text{fit}}-\pt^{i, \text{meas}})^{2}}{{\sigma_{i}}^{2}}
+ \sum_{j= x,\ y} \frac{({p_{j}}^{\mathrm{UE},\text{fit}}-{p_{j}}^{\mathrm{UE},\text{meas}})^{2} }{{\sigma_{\mathrm{UE}}}^{2}}  \\
+ \frac{(M_{\ell\nu}-M_{\PW})^{2}}{{\Gamma_{\PW}}^{2}}
+ \sum_{k= {\cPqt^\text{had},\ \cPqt^\text{lep}}} \frac{(M_k-M_{\cPqt})^{2}}{{\Gamma_{\cPqt}}^{2}} \Biggr).
\label{eq:chi2}
\end{multline}
\end{linenomath}
In the first two terms, the momentum with superscript ``fit'' is the variable to be determined by the fit, and the measured TS-corrected input $\pt$ is denoted with the superscript ``meas''. The first term fits the transverse momentum of the lepton and leading four jets and the second term fits an unclustered energy (UE) in the transverse directions $x$ and $y$.
The unclustered transverse energy vector is obtained from all the observables in the transverse plane by the relation:
\begin{linenomath}
\begin{equation}\label{eq:met}
p_{x, y}^{\mathrm{UE}} = - \sum_{i=\ell,\ 4\text{jets}} p_{x, y}^{i} - \sum_{j= \text{extra jets},\ \pt>10\GeV,\ \abs{\eta}<2.5} p_{x, y}^{j}  - p^{\text{miss}}_{x, y},
\end{equation}
\end{linenomath}
where the $p^{\text{miss}}_{x}$  and $p^{\text{miss}}_{y}$ are the $x$ and $y$ components of  $\ptvecmiss$.
Variation of the lepton, jet, and UE is allowed within the measurement uncertainties, $\sigma_{i}$ and $\sigma_{\mathrm{UE}}$, depending on their $\pt$.  The longitudinal momentum ($\pzn$) of the neutrino is calculated by the leptonic ($\ell \nu$) $\PW$ boson mass constraint ($[\mathbf{p}^{\ell}+\mathbf{p}^{\nu}]^2=M^{2}_{\PW}$) and only real $\pzn$ is taken into account in the fit.  During the iterations for minimizing the $\chi^2$, this $\pzn$ varies to keep the $\PW$ boson mass constrained. The neutrino momentum vector ($p_{x}^{\Pgn, \text{fit}}$, $p_{y}^{\Pgn, \text{fit}}$,  $p_{z}^{\Pgn, \text{fit}}$) is reconstructed from all the fitted momenta and Eq.~\ref{eq:met}: $p_{x, y}^{\Pgn, \text{fit}} = p_{x, y}^{\text{miss},\text{fit}}$.
 The last term constrains the hadronic and leptonic top quark candidates to have the true mass of 172.5\GeV. The widths of the $\PW$ boson ($\Gamma_{\PW}$) and top quark ($\Gamma_{\cPqt}$)  in Ref.~\cite{pdg} are used for the resolution in the fit.
The $\chi^{2}$ minimization is performed for each possible combination of the four leading jets to quarks in the $\ttbar$ system, where the $\cPqb$-tagged jets are only assigned to the $\cPqb$ quark daughters.
In order to suppress combinatorial backgrounds and the irreducible contaminations from initial- and final-state radiation jets, two requirements are imposed:
$|\pt^{\text{jet, meas}} - \pt^{\text{jet, fit}}| <  20\GeV$ for the jets used in the fit and $M_{k} < 200\GeV$, in which $M_{k}$ is reconstructed using input jets before the $\chi^{2}$ fit, for the hadronically decaying top quark.
In the jet-quark assignment that minimizes the $\chi^{2}$, the two jets not assigned to either $\cPqb$ quarks originating directly from top quark decays form a $\ch \to \cb$ candidate.

The reconstructed events are further categorized according to the lepton flavour ($\Pgm$ or $\Pe$) and the number of $\cPqb$-tagged jets (2 or $\geq$3). Events containing two $\cPqb$-tagged jets are used to constrain the SM $\ttbar$ background, while events with three or more $\cPqb$-tagged jets are used to search directly the presence of $\hcb$ decays.
In events with two $\cPqb$ tags, the fit has only two possible combinations of the jet assignment. However, in events with three or more $\cPqb$ tags, one $\cPqb$-tagged jet is assigned to a leptonically decaying top quark, and two other $\cPqb$-tagged jets are assigned to the hadronically decaying top quark resulting in additional ambiguity.
According to simulation, the ambiguity is efficiently resolved by the fit procedure only for $\ch$ masses below 120\GeV. At higher masses (130--150\GeV), the ambiguity is resolved by assigning the $\cPqb$ jet with the lower $\pt$ to the $\cPqb$ quark that originates from the $\tchb$ decay.

\section{Systematic uncertainties} \label{sec:sys}

\begin{table}[bp]
\topcaption{Summary of the relative systematic uncertainties in the event yields for the $\ch$ signal ($\mh=120\GeV$), simulated SM backgrounds (separated into $\ttbar$ and non-$\ttbar$ components), and the data-driven multijet events. The uncertainties apply to both $\Pgm$+jets and $\Pe$+jets events, and in the case where the uncertainties in the two channels differ, a range is given. Uncertainties on the shape of templates are marked with an asterisk.}\label{tab:sys}
\centering
\cmsTable{
\begin{tabular}{l|cc|cc|cc}
\hline
\multirow{2}{*}{Source of uncertainty}   &   \multicolumn{ 2}{c|}{Signal ($\mh$ = 120\GeV) (\%)} &  \multicolumn{ 2}{c|}{\ttbar (\%)} &  \multicolumn{ 2}{c}{Non-\ttbar (\%)}  \Tstrut\\
                        &  2 $\cPqb$ tags   & $\geq$ 3 $\cPqb$ tags &  2 $\cPqb$ tags   & $\geq$3 $\cPqb$ tags &    2 $\cPqb$ tags   & $\geq$3 $\cPqb$ tags \Bstrut\\
\hline
Jet energy scale (JES)*  & 4.6--5.3 & 5.0--5.8 &  3.1--3.3 & 3.1  &  10.2--14.5  & 1.9--3.4     \Tstrut\\
Flavour-dependent JES ($\cPqb$ quark)*  & 0.4 & 0.5  & 0.1 &  0.1  & 0.2 & 0.5--3.4       \\
Flavour-dependent JES (udsc quark or gluon)*  & 1.0  & 0.4 & 0.9  & 0.8  & 2.8--4.6 & 2.7--9.0   \\
Jet energy resolution*  & 0.2 & 0.8  & 0.3  & 0.3  & 1.0--1.3  &  1.3--4.9    \Bstrut\\[\cmsTabSkip]
$\cPqb$ tagging scale factor for b/c-quark jets*  & 1.2   & 5.7  & 3.6 & 5.7   & 0.6--0.8 & 2.0--3.8    \Tstrut\\
Mistag scale factor for light quark jets*  & 0.2  & 0.3  & 0.2 & 2.7  & 0.9--1.5  & 0.9--2.0   \Bstrut\\[\cmsTabSkip]

\ttbar \pt reweighting*  & 0.2  & 1.0  & 1.4--1.7  &  1.6--1.9 &  \NA & \NA \Tstrut\\
NLO-vs.-LO shape*   &  7.5--8.4 & 7.2--7.7 & 7.0--8.2 & 6.8--7.6  & \NA  & \NA  \\
ME-PS matching*     & 0.8 & 0.9 & 1.1 & 1.8--2.4 & \NA & \NA \\
Renormalization and factorization scales*  & 0.3 & 1.3--1.8  & 0.8--1.8  & 1.3--1.6 & \NA  & \NA      \\
Top quark mass*  & 1.1--1.4 & 1.1--1.5 & 0.4--1.2 & 0.9 & \NA & \NA \\
$\ttbb$ production rescaling*   & \NA & \NA & 3.7--3.9 & 10.2--10.9 & \NA  & \NA \\
\PYTHIA--\MADGRAPH $\pt(\ttbar)$ difference*  & 0.1 & 0.1 & \NA  & \NA & \NA & \NA \Bstrut \\[\cmsTabSkip]

\ttbar cross section &  6.5 & 6.5 &  6.5 &  6.5 & \NA & \NA \Tstrut \\
Integrated luminosity  & \multicolumn{ 6}{c}{2.6}  \\
Muon scale factor ($\Pgm$+jets)  & \multicolumn{ 6}{c}{3.0} \\
Electron scale factor ($\Pe$+jets)  &\multicolumn{ 6}{c}{3.0}  \\
Pileup reweighting  &   \multicolumn{ 6}{c}{0.1--1.3}   \Bstrut\\[\cmsTabSkip]
Multijet background prediction from data* &  \multicolumn{ 4}{c}{\NA} & 0.3--2.3 & 5.2--10.7 \TBstrut\\
\hline
\end{tabular}
}
\end{table}

Systematic uncertainties can affect the overall signal and background events, as well as cause distortions in the shape of the dijet mass distribution. Since the $\ch$ originates from a top quark decay, a number of systematic uncertainties in the $\ch$ signal and SM $\ttbar$ background are correlated. The systematic uncertainties are estimated based on the samples and methods used in Ref.~\cite{Khachatryan:2015hba}. A summary of the systematic uncertainties is given in Table~\ref{tab:sys}.

Sources of systematic uncertainties are grouped into several categories: jet corrections, $\cPqb$ tagging effects, $\ttbar$ modeling, and normalizations.  Uncertainties due to jet energy corrections, flavour-dependent uncertainties, and
uncertainties due to jet energy resolution corrections are estimated by varying the correction factors by $\pm$1 standard deviation (s.d.). The efficiency difference from data to the simulation (scale factor) in heavy quark tagging ($\cPqb/\cPqc$ jets) and mistagging for light-flavoured jets is also varied by $\pm$1 s.d. separately and the corresponding changes are estimated. Similarly, the following quantities are also varied by $\pm$1 s.d.: normalization of the $\ttbar$ cross section in the simulation, integrated luminosity~\cite{lumi_unc} of the data sample, and lepton scale factors including the single-lepton trigger, identification, and relative isolation.
The uncertainty due to pileup is estimated by varying the total inelastic cross section used in the simulation by $\pm$5\%~\cite{CMS:2014yxa}.

To account for the uncertainties in the modeling of SM \ttbar events, we consider the uncertainty in reweighting the shape of the top quark \pt distribution in the \ttbar events  to match the simulation to data, NLO production versus LO production with 0--3 partons ({\POWHEG} versus {\MADGRAPH}),
matching thresholds used for interfacing the matrix-elements calculations of the \MADGRAPH generator to the \PYTHIA parton showers (ME-PS), renormalization and factorization scales, and the uncertainty in the top quark mass of $172.5 \pm 1.0\GeV$. The uncertainty in the $\ttbb$ rescaling ratio is estimated to be 50\%, combining the $\ttbb$ cross section uncertainties (42\%) and a few percent of  the inefficiency of counting $\cPqb$ jets in generator level. The $\ttbb$ rescaling uncertainties listed in Table~\ref{tab:sys} are the impact of rescaling on the selected $\ttbar$ events.

The systematic uncertainty in the SM \ttbar modeling is estimated using simulation samples in which the corresponding systematic sources are varied.
In order to estimate the \ttbar modeling uncertainties in the simulated $\ch$ signal events, the $\pt$ distribution of the top quarks from SM \ttbar events is used. The ratio of the $\pt$ distribution with each parameter shifted to the nominal value is calculated, then is used to reweight the top quark $\pt$ distributions in the \ch signal simulation to mimic the systematic sample.  By using this method the modeling uncertainties for $\ch$ signal events are estimated as listed in Table~\ref{tab:sys}.
In addition, as the \ch events are generated using \PYTHIA, the difference in \ttbar generation estimated by the top quark \pt distributions of \PYTHIA and \MADGRAPH, is then used as an additional systematic uncertainty.

\section{Results}

\begin{figure}[htbp]
\centering
    \includegraphics[width=0.40\textwidth]{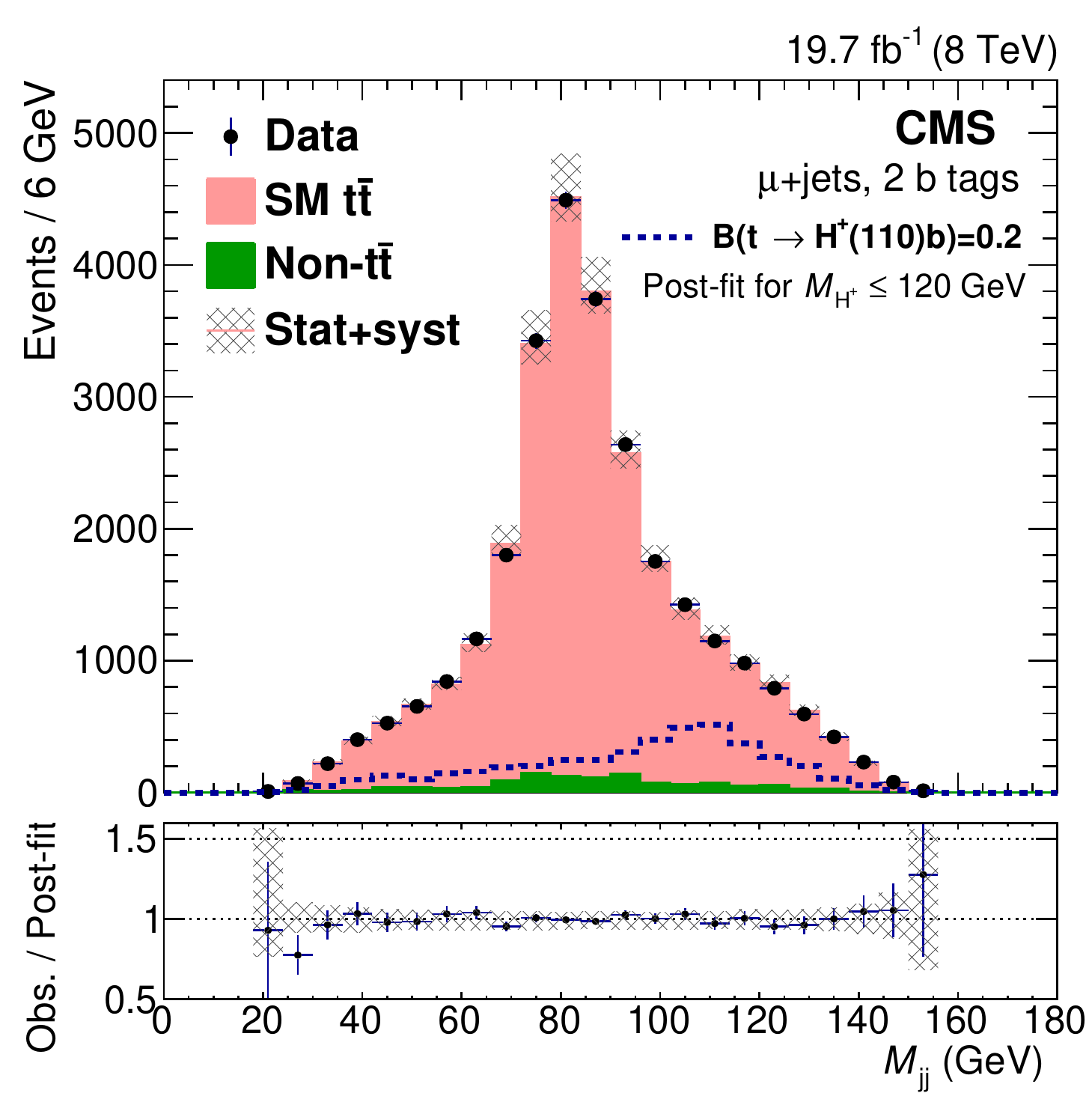}
    \includegraphics[width=0.40\textwidth]{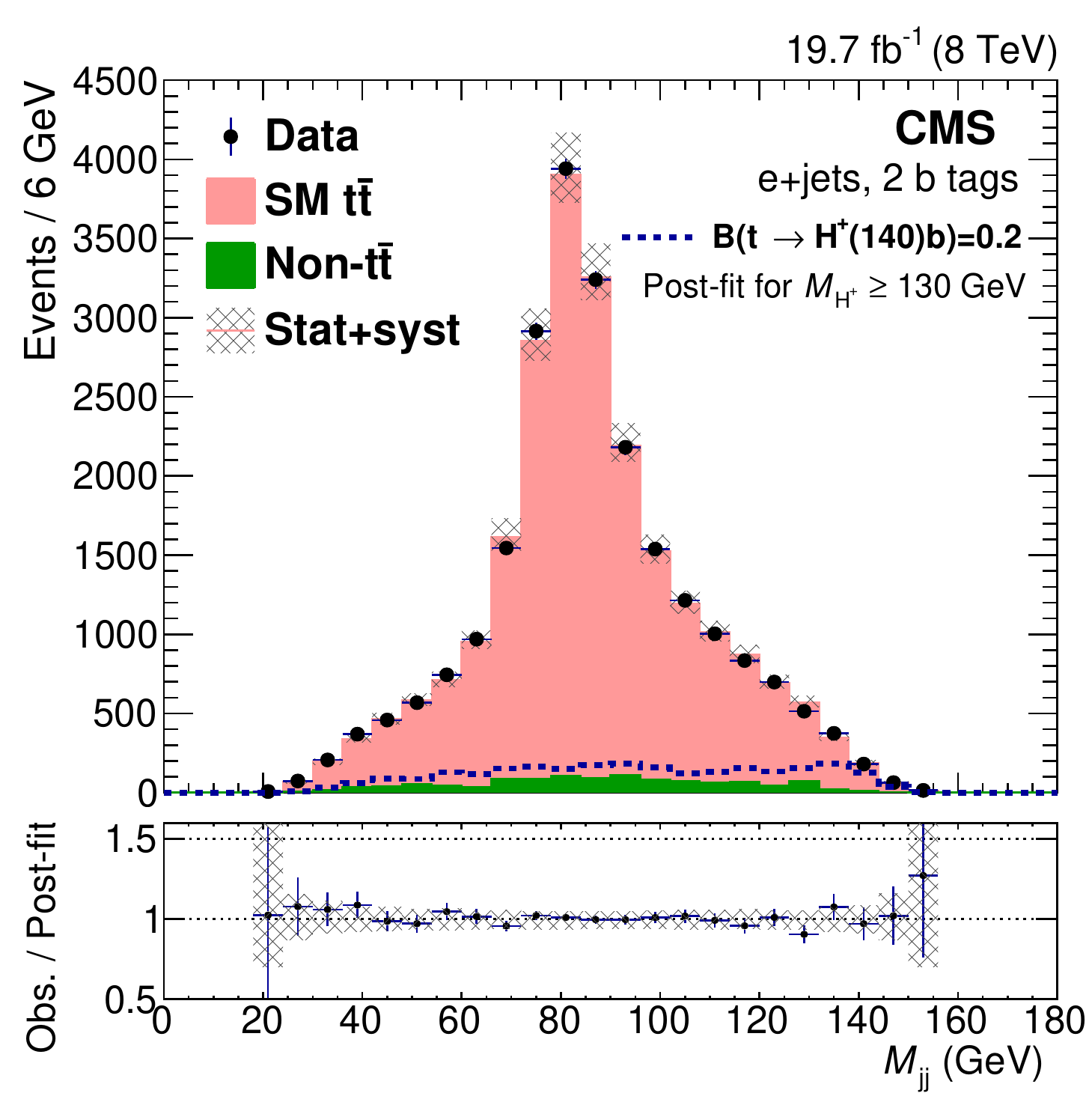}

    \includegraphics[width=0.40\textwidth]{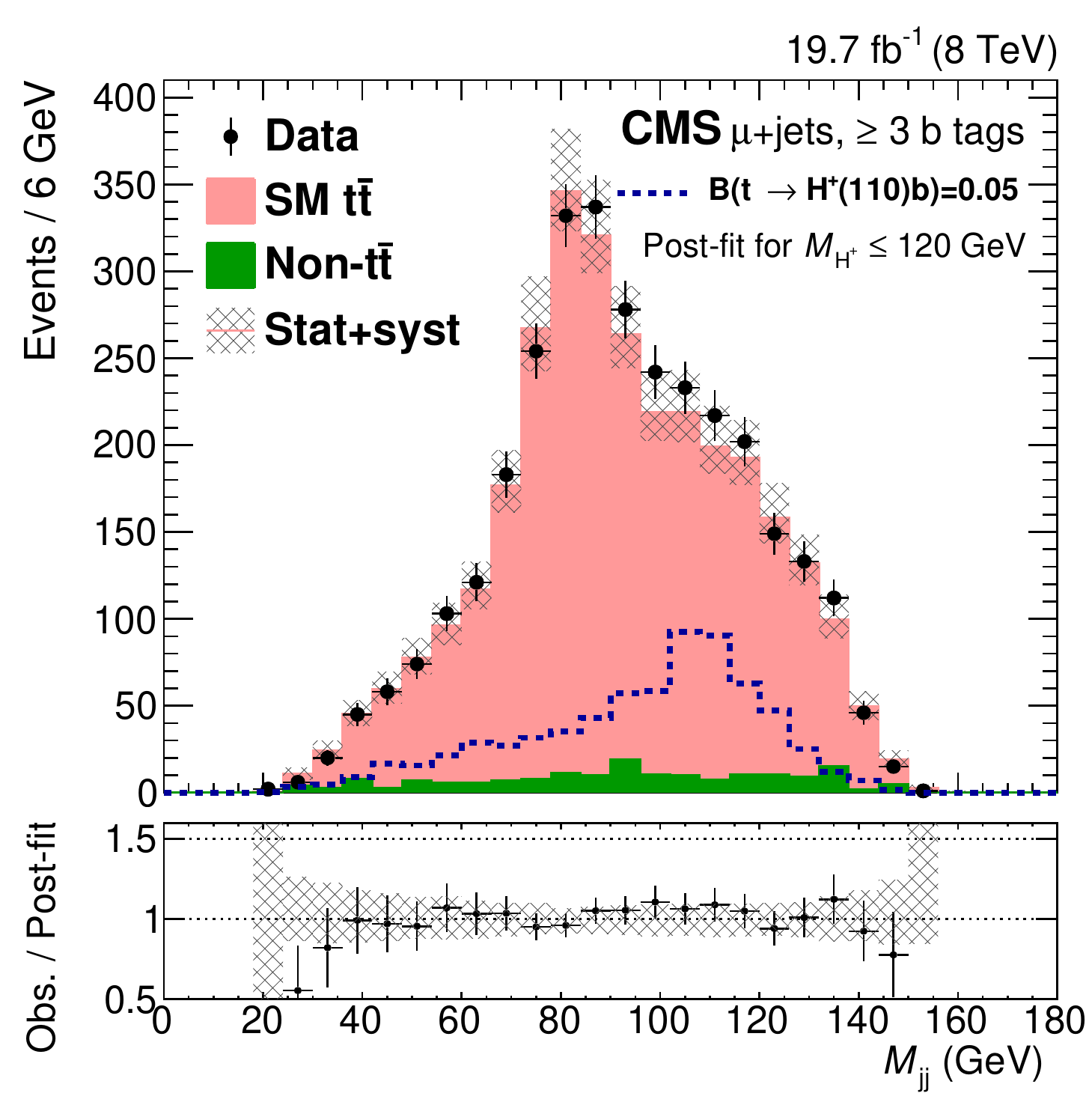}
    \includegraphics[width=0.40\textwidth]{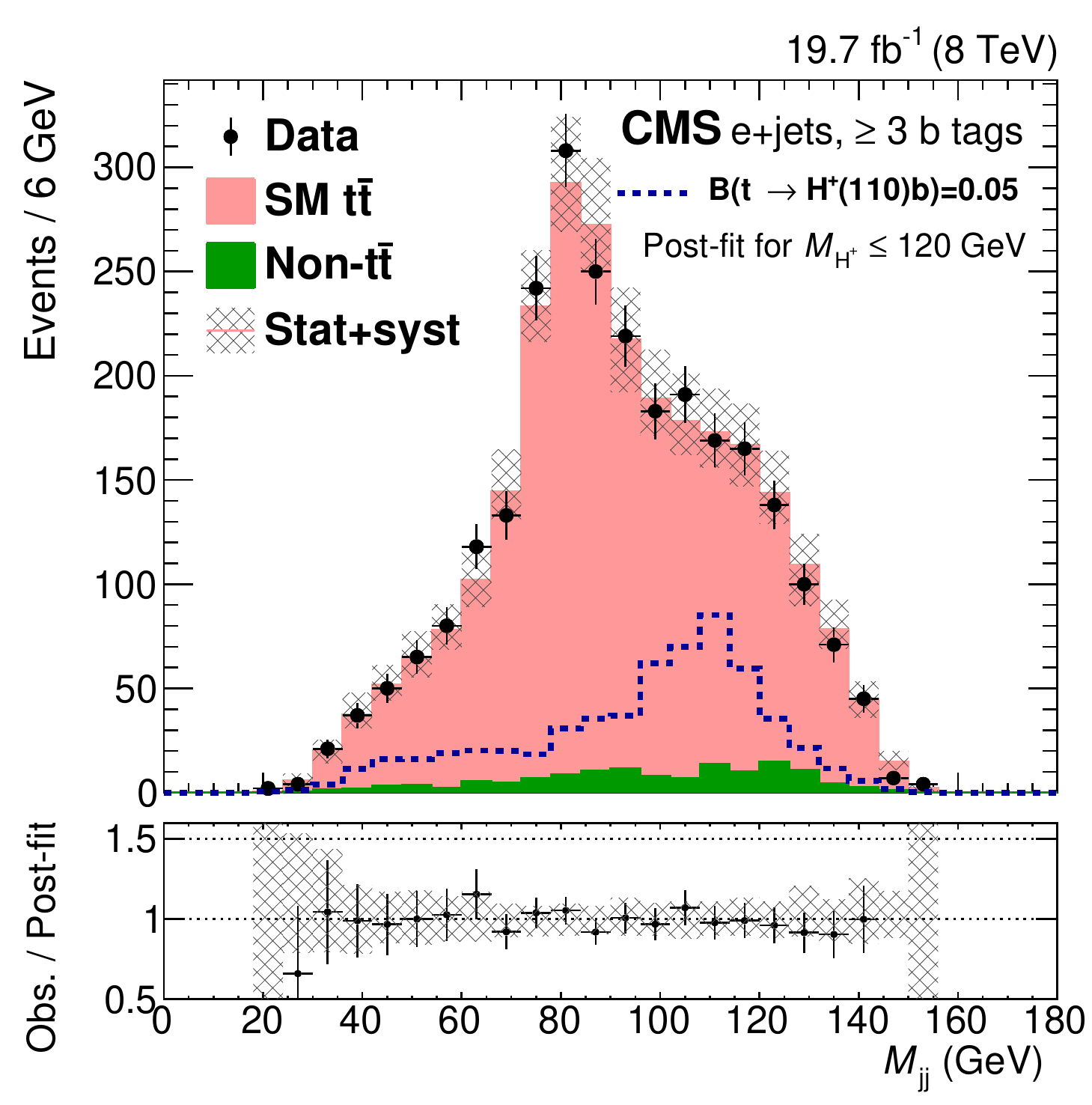}

    \includegraphics[width=0.40\textwidth]{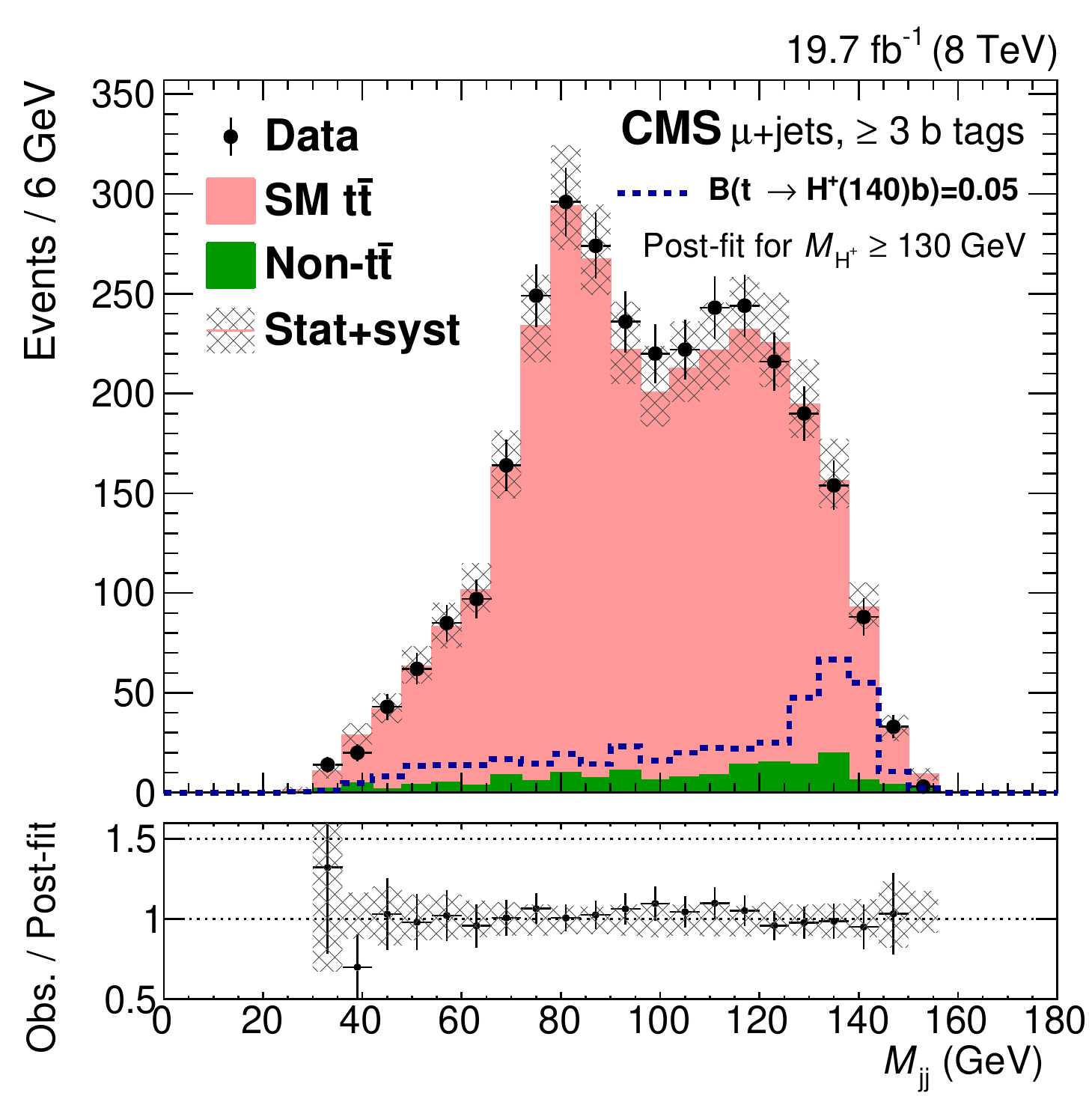}
    \includegraphics[width=0.40\textwidth]{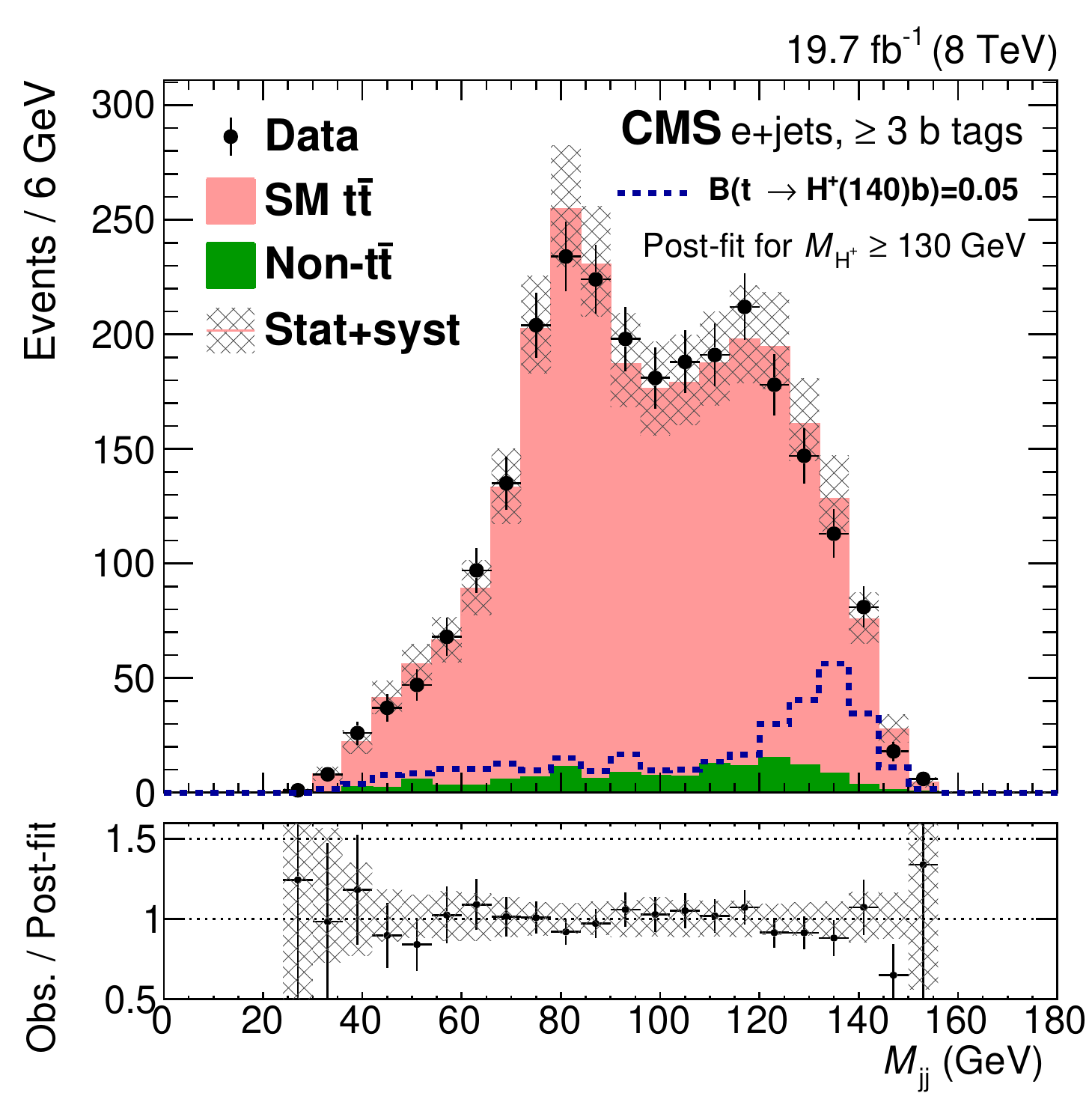}
    \caption{Post-fit with a null-$\ch$ hypothesis on the expected dijet mass distributions from SM backgrounds (cumulative filled histograms) and their ratio of observed to predicted yields for the $\Pgm$+jets (left column) and $\Pe$+jets (right column) channels. In the first row, events are shown for two $\cPqb$ tags together with the fit procedure for a $\ch$ signal ($\mh$ = 110\GeV in left and 140\GeV in right). The second (third) row shows the results for events with at least three $\cPqb$ tags in the fit procedure for the $\ch$ search with $\mh$ = 90--120 (130--150)\GeV.  The dijet distributions are compared with the $\ch$ signal shape (dashed line) for $\mh = 110$ and 140\GeV.}\label{fig:mlf}
\end{figure}

\begin{figure}[hbtp]
\centering%
    \includegraphics[width=0.85\textwidth]{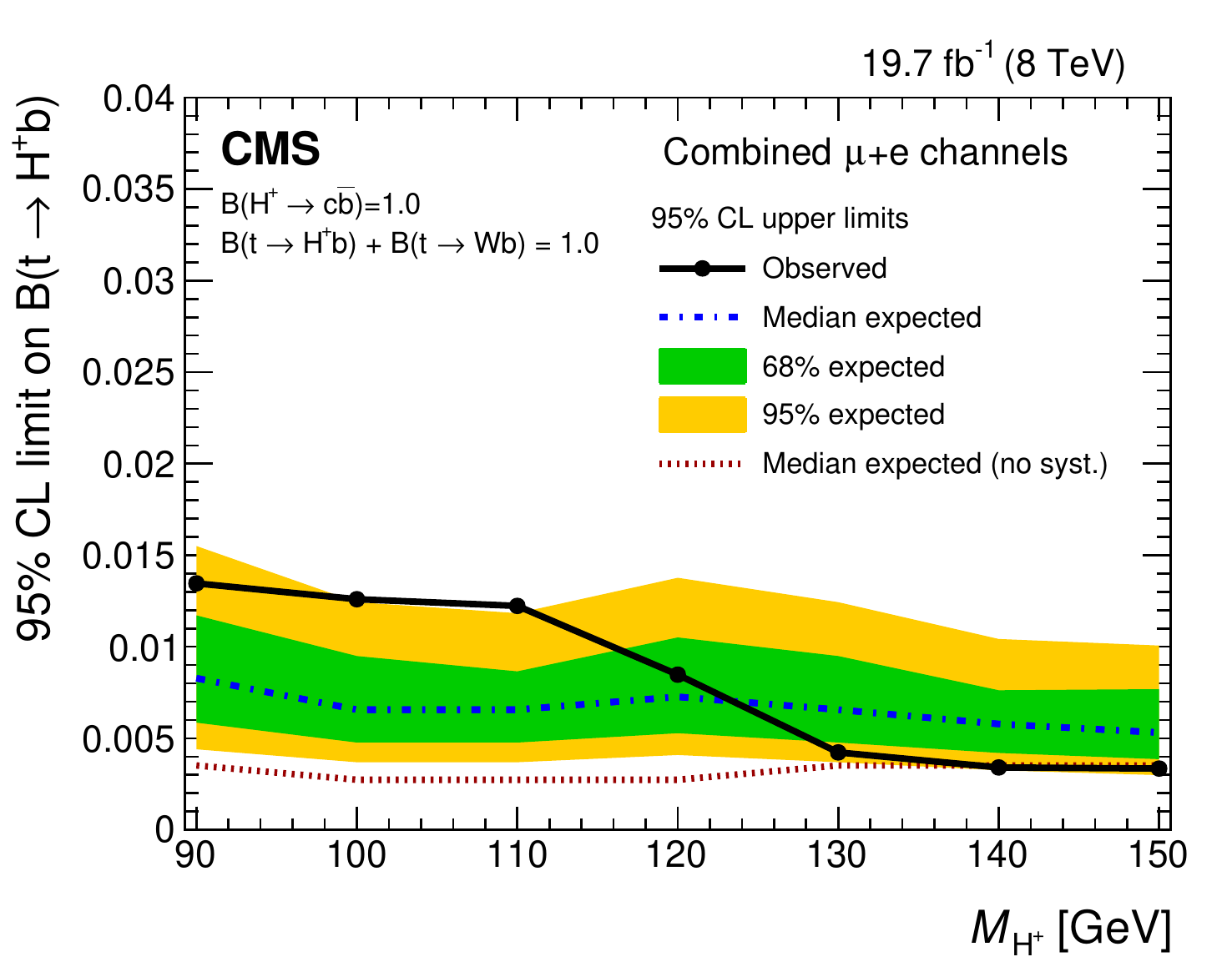}
    \caption{Upper limits at the 95\% confidence level ($\CL$) on the branching fraction $\br(\tchb)$, assuming $\br(\hcb) = 1.0$ and $\br(\tchb) + \br(\twb) = 1.0$, for the combined $\Pgm$+jets and $\Pe$+jets channels. The black solid line shows the observed limit. The mean expected limit is shown as a blue dashed line and the green/yellow bands indicate the 68/95\% confidence intervals for the expected limits. The red dotted line shows the mean expected limit in the absence of systematic uncertainties.}
    \label{fig:limit_final_comb}
\end{figure}

Figure~\ref{fig:mlf} shows the dijet mass distributions together with the expected SM processes and $\ch$ signal after the  kinematic fit procedures in {\Pgm}+jets and {\Pe}+jets events with two $\cPqb$ tags and at least three $\cPqb$ tags, which are used for the $\ch$ search with $\mh$ of 90--120 and 130--150\GeV.
A binned maximum likelihood fit is performed simultaneously to all the observed dijet mass distributions, using the signal and background templates extracted from the simulation or from the data. The background templates are composed of the dominant SM $\ttbar$ and non-$\ttbar$ contributions.
For the $\mh$ values of 120 and 130\GeV, where the kinematic fit procedure changes as described in Section~\ref{rec_tt}, the limits are derived also with the alternate procedure, giving consistent results.
No significant excess is seen above the expected SM background.  The upper limits at 95\% \CL on the branching fraction $\br(\tchb)$ are calculated using the statistical tools in \textsc{RooStat}~\cite{ref:combine} and the \CL{}$_\mathrm{s}$ criterion~\cite{CLS1,CLS2} with a profile likelihood ratio as a test statistic~\cite{CLs} and using an asymptotic formulae~\cite{Cowan:2010js}.
The expected branching fraction limit is calculated using an Asimov dataset with a null hypothesis.
Systematic uncertainties are treated as nuisance parameters and profiled in the fit following a log-normal distribution for the normalization uncertainties and using distorted templates for shape systematic uncertainties.
With the assumptions of $\br(\hcb) = 1.0$ and $\br(\tchb)+\br(\twb) = 1.0$, the expected and observed limits as a function of $\mh$ are shown in Fig.~\ref{fig:limit_final_comb}.
The expected limits without systematic uncertainties are also shown to illustrate that the analysis sensitivity is largely limited by the present level of our knowledge of the systematic uncertainties. The biggest impact on the expected limit comes from the $\ttbb$ production rescaling uncertainty.

\section{Summary}

A search for  charged Higgs boson decaying to a charm and a bottom quark ($\hcb$) is performed for the first time.
The search uses \ttbar events with a final state containing at least four jets, a charged lepton (muon or electron), and missing transverse momentum.  The search is based on the analysis of proton-proton collision data recorded at $\sqrt{s} = 8\TeV$, corresponding to an integrated luminosity of 19.7\fbinv.
A kinematic fit is performed to identify the pair of jets least likely to be the $\cPqb$ quarks originating from direct top quark decays and the invariant mass of this pair is used as the final observable in the search. No evidence for the presence of a charged Higgs boson is observed and upper limits at 95\% confidence level of 0.8--0.5\% are set on the branching fraction $\br(\tchb)$, assuming $\br(\hcb) = 1.0$ and $\br(\tchb) + \br(\twb) = 1.0$,  for the charged Higgs boson mass range 90--150\GeV.

\begin{acknowledgments}
\hyphenation{Bundes-ministerium Forschungs-gemeinschaft Forschungs-zentren Rachada-pisek} We congratulate our colleagues in the CERN accelerator departments for the excellent performance of the LHC and thank the technical and administrative staffs at CERN and at other CMS institutes for their contributions to the success of the CMS effort. In addition, we gratefully acknowledge the computing centres and personnel of the Worldwide LHC Computing Grid for delivering so effectively the computing infrastructure essential to our analyses. Finally, we acknowledge the enduring support for the construction and operation of the LHC and the CMS detector provided by the following funding agencies: the Austrian Federal Ministry of Science, Research and Economy and the Austrian Science Fund; the Belgian Fonds de la Recherche Scientifique, and Fonds voor Wetenschappelijk Onderzoek; the Brazilian Funding Agencies (CNPq, CAPES, FAPERJ, and FAPESP); the Bulgarian Ministry of Education and Science; CERN; the Chinese Academy of Sciences, Ministry of Science and Technology, and National Natural Science Foundation of China; the Colombian Funding Agency (COLCIENCIAS); the Croatian Ministry of Science, Education and Sport, and the Croatian Science Foundation; the Research Promotion Foundation, Cyprus; the Secretariat for Higher Education, Science, Technology and Innovation, Ecuador; the Ministry of Education and Research, Estonian Research Council via IUT23-4 and IUT23-6 and European Regional Development Fund, Estonia; the Academy of Finland, Finnish Ministry of Education and Culture, and Helsinki Institute of Physics; the Institut National de Physique Nucl\'eaire et de Physique des Particules~/~CNRS, and Commissariat \`a l'\'Energie Atomique et aux \'Energies Alternatives~/~CEA, France; the Bundesministerium f\"ur Bildung und Forschung, Deutsche Forschungsgemeinschaft, and Helmholtz-Gemeinschaft Deutscher Forschungszentren, Germany; the General Secretariat for Research and Technology, Greece; the National Research, Development and Innovation Fund, Hungary; the Department of Atomic Energy and the Department of Science and Technology, India; the Institute for Studies in Theoretical Physics and Mathematics, Iran; the Science Foundation, Ireland; the Istituto Nazionale di Fisica Nucleare, Italy; the Ministry of Science, ICT and Future Planning, and National Research Foundation (NRF), Republic of Korea; the Lithuanian Academy of Sciences; the Ministry of Education, and University of Malaya (Malaysia); the Mexican Funding Agencies (BUAP, CINVESTAV, CONACYT, LNS, SEP, and UASLP-FAI); the Ministry of Business, Innovation and Employment, New Zealand; the Pakistan Atomic Energy Commission; the Ministry of Science and Higher Education and the National Science Centre, Poland; the Funda\c{c}\~ao para a Ci\^encia e a Tecnologia, Portugal; JINR, Dubna; the Ministry of Education and Science of the Russian Federation, the Federal Agency of Atomic Energy of the Russian Federation, Russian Academy of Sciences, the Russian Foundation for Basic Research and the Russian Competitiveness Program of NRNU ``MEPhI"; the Ministry of Education, Science and Technological Development of Serbia; the Secretar\'{\i}a de Estado de Investigaci\'on, Desarrollo e Innovaci\'on, Programa Consolider-Ingenio 2010, Plan Estatal de Investigaci\'on Cient\'{\i}fica y T\'ecnica y de Innovaci\'on 2013-2016, Plan de Ciencia, Tecnolog\'{i}a e Innovaci\'on 2013-2017 del Principado de Asturias and Fondo Europeo de Desarrollo Regional, Spain; the Swiss Funding Agencies (ETH Board, ETH Zurich, PSI, SNF, UniZH, Canton Zurich, and SER); the Ministry of Science and Technology, Taipei; the Thailand Center of Excellence in Physics, the Institute for the Promotion of Teaching Science and Technology of Thailand, Special Task Force for Activating Research and the National Science and Technology Development Agency of Thailand; the Scientific and Technical Research Council of Turkey, and Turkish Atomic Energy Authority; the National Academy of Sciences of Ukraine, and State Fund for Fundamental Researches, Ukraine; the Science and Technology Facilities Council, UK; the US Department of Energy, and the US National Science Foundation.

Individuals have received support from the Marie-Curie programme and the European Research Council and Horizon 2020 Grant, contract No. 675440 (European Union); the Leventis Foundation; the A. P. Sloan Foundation; the Alexander von Humboldt Foundation; the Belgian Federal Science Policy Office; the Fonds pour la Formation \`a la Recherche dans l'Industrie et dans l'Agriculture (FRIA-Belgium); the Agentschap voor Innovatie door Wetenschap en Technologie (IWT-Belgium); the F.R.S.-FNRS and FWO (Belgium) under the ``Excellence of Science - EOS" - be.h project n. 30820817; the Ministry of Education, Youth and Sports (MEYS) of the Czech Republic; the Lend\"ulet (``Momentum") Programme and the J\'anos Bolyai Research Scholarship of the Hungarian Academy of Sciences, the New National Excellence Program \'UNKP, the NKFIA research grants 123842, 123959, 124845, 124850 and 125105 (Hungary); the Council of Scientific and Industrial Research, India; the HOMING PLUS programme of the Foundation for Polish Science, cofinanced from European Union, Regional Development Fund, the Mobility Plus programme of the Ministry of Science and Higher Education, the National Science Center (Poland), contracts Harmonia 2014/14/M/ST2/00428, Opus 2014/13/B/ST2/02543, 2014/15/B/ST2/03998, and 2015/19/B/ST2/02861, Sonata-bis 2012/07/E/ST2/01406; the National Priorities Research Program by Qatar National Research Fund; the Programa de Excelencia Mar\'{i}a de Maeztu and the Programa Severo Ochoa del Principado de Asturias; the Thalis and Aristeia programmes cofinanced by EU-ESF and the Greek NSRF; the Rachadapisek Sompot Fund for Postdoctoral Fellowship, Chulalongkorn University and the Chulalongkorn Academic into Its 2nd Century Project Advancement Project (Thailand); the Welch Foundation, contract C-1845; and the Weston Havens Foundation (USA).

\end{acknowledgments}

\bibliography{auto_generated}

\cleardoublepage \appendix\section{The CMS Collaboration \label{app:collab}}\begin{sloppypar}\hyphenpenalty=5000\widowpenalty=500\clubpenalty=5000\vskip\cmsinstskip
\textbf{Yerevan Physics Institute, Yerevan, Armenia}\\*[0pt]
A.M.~Sirunyan, A.~Tumasyan
\vskip\cmsinstskip
\textbf{Institut f\"{u}r Hochenergiephysik, Wien, Austria}\\*[0pt]
W.~Adam, F.~Ambrogi, E.~Asilar, T.~Bergauer, J.~Brandstetter, E.~Brondolin, M.~Dragicevic, J.~Er\"{o}, A.~Escalante~Del~Valle, M.~Flechl, R.~Fr\"{u}hwirth\cmsAuthorMark{1}, V.M.~Ghete, J.~Hrubec, M.~Jeitler\cmsAuthorMark{1}, N.~Krammer, I.~Kr\"{a}tschmer, D.~Liko, T.~Madlener, I.~Mikulec, N.~Rad, H.~Rohringer, J.~Schieck\cmsAuthorMark{1}, R.~Sch\"{o}fbeck, M.~Spanring, D.~Spitzbart, A.~Taurok, W.~Waltenberger, J.~Wittmann, C.-E.~Wulz\cmsAuthorMark{1}, M.~Zarucki
% ...
\vskip\cmsinstskip
% ...
1:  Also at Vienna University of Technology, Vienna, Austria\\
% ...
\end{sloppypar}
\end{document}